\documentclass[english,aps,showpacs]{revtex4}
\usepackage[T1]{fontenc}
\usepackage[latin9]{inputenc}
\usepackage{float}
\usepackage{amsthm}
\usepackage{amsmath}
\usepackage{amssymb}

\makeatletter

\AtBeginDocument{}

\DeclareFontEncoding{LGR}{}{}

\providecommand{\tabularnewline}{\\}

\@ifundefined{textcolor}{}
{%
 \definecolor{BLACK}{gray}{0}
 \definecolor{WHITE}{gray}{1}
 \definecolor{RED}{rgb}{1,0,0}
 \definecolor{GREEN}{rgb}{0,1,0}
 \definecolor{BLUE}{rgb}{0,0,1}
 \definecolor{CYAN}{cmyk}{1,0,0,0}
 \definecolor{MAGENTA}{cmyk}{0,1,0,0}
 \definecolor{YELLOW}{cmyk}{0,0,1,0}
 }

\makeatother

\usepackage{babel}

\begin{document}

\title{Probing the Effects of Lorentz-Symmetry Violating Chern-Simons and
Ricci-Cotton Terms in Higher Derivative Gravity}

\author{B. Pereira-Dias}

\email{bpdias@cbpf.br}

\affiliation{Centro Brasileiro de Pesquisas Físicas, Rua Dr. Xavier Sigaud 150,
Urca,\\
 Rio de Janeiro, Brazil, CEP 22290-180}

\author{C. A. Hernaski}

\email{carlos@cbpf.br}

\affiliation{Centro Brasileiro de Pesquisas Físicas, Rua Dr. Xavier Sigaud 150,
Urca,\\
 Rio de Janeiro, Brazil, CEP 22290-180}

\author{J. A. Helayël-Neto}

\email{helayel@cbpf.br}

\affiliation{Centro Brasileiro de Pesquisas Físicas, Rua Dr. Xavier Sigaud 150,
Urca,\\
 Rio de Janeiro, Brazil, CEP 22290-180}

\begin{abstract}
The combined effects of the Lorentz-symmetry violating Chern-Simons
and Ricci-Cotton actions are investigated for the Einstein-Hilbert
gravity in the second order formalism modified by higher-derivative
terms, and their consequences on the spectrum of excitations are analyzed.
We follow the lines of previous works and build up an orthonormal
basis of projector-like operators for the degrees of freedom, rather
than for the spin modes of the fields. With this new basis, the attainment
of the propagators is remarkably simplified and the identification
of the physical and unphysical modes becomes more immediate.  Our
conclusion is that the only tachyon- and ghost-free model is the Einstein-Hilbert
action added up by the Chern-Simons term with a time-like vector of
the type $v^{\mu}=\left(\mu,\vec{0}\right)$. Spectral consistency
imposes that the Ricci-Cotton term must be switched off. We then infer
that gravity with Lorentz-symmetry violation imposes a drastically
different constraint on the background if compared to ordinary gauge
theories whenever conditions for the suppression of tachyons and ghosts
are imposed.

\end{abstract}
\maketitle

\section{Introduction}

In spite of the special role that Lorentz symmetry plays in fundamental
Particle Physics, over the last two decades there has been a remarkable
activity in considering models where this symmetry is violated. One
reason for this excitement is that the main candidates for a consistent
quantum theory of gravitation, such as loop quantum gravity \cite{Ashtekar,Rovelli}
and Horava-Lifshitz gravity \cite{Horava}, exhibit a phase where
Lorentz symmetry is broken; on the other hand, string theory may spontaneously
break Lorentz invariance by the vacuum condensation of non-trivial
Lorentz tensors \cite{KostSamu1989}. Most interestingly, the relics
of the Lorentz-symmetry violation (LV) could be detectable at low-energy
experimental measurements, yielding physical constraints on these
fundamental theories.

The search for extended quantum gravity models was mainly dictated
by the difficulty of obtaining, simultaneously, a renormalizable and
unitary quantum field theory to describe gravity. The attempts to
simply add relativistic higher-derivative corrections to the Einstein-Hilbert
Lagrangian did not solve such problem. Although these terms improve
the ultraviolet divergences, they may also yield ghost excitations
which jeopardizes unitarity of S-matrix. (For a deeper discussion,
we address the reader to the works of Refs. \cite{Stell1979,Stelle1978,Deser1974}.)
With this in mind, we judge that a first and important test for the
quantum consistency of any modified gravity theory is to require unitarity
in the sense that its particle spectrum does not propagate neither
tachyon nor ghost modes. It is important to clearly state that, in
this paper, we shall adopt a point of view that the models of gravity
we are dealing with are understood as effective field theories. Their
ultraviolet completion must come from a more fundamental theory which
is expected to be renormalizable, such as string or M-theory.

The type of gravity theory we shall inspect in this paper is of the
LV-type characterized by the presence of a constant vector, $v^{a}$,
that spoils the isotropy of space-time. In \cite{JackiwPi}, this
vector is coupled to gravitation via a Chern-Simons-like term. A great
motivation for considering the 4D C-S-like term comes from the striking
effect of the Chern-Simons (C-S) term in 3-D that yields a consistent
description of a massive graviton. In this paper we intend to verify
whether this consistent mass generating mechanism survives in 4D.
Furthermore, along with the Einstein-Hilbert and C-S term, we also
consider the effects of (curvature)$^{2}$-terms and an extra LV-term,
built up in analogy with the Ricci-Cotton (R-C) term in 3-D. A broad
review on LV CS gravity is given in \cite{AlexYunesRept}. For further
discussion see \cite{Ferrari,Furtado,Ahmedov,Ertem,Li,Alexander,Cantcheff,Nojiri1,Nojiri2,Ghodsi}.

Invoking spatial isotropy, the authors of \cite{JackiwPi} restrict
their discussion to $v^{a}=\left(\mu,\vec{0}\right)$. For the sake
of generality, and motivated by recent works that point to a possible
spatial anisotropy at cosmological scales \cite{Webb2010,KostMewes},
we shall not restrict our discussion to any particular choice of $v^{a}$.
In fact, it is known that the nature of the LV background vector may
drastically change the spectrum of the model. For example, in the
electromagnetic C-S LV model, it has been argued that a space-like
Lorentz background vector of the type $v^{a}=\left(0,\vec{\mu}\right)$
renders the theory free from ghosts and tachyons, whereas a time-like
vector of the type $v^{a}=\left(\mu,\vec{0}\right)$ yields an inconsistent
quantum theory \cite{Adam-Klinkhamer,Scarpelli-Belich}. In this paper,
surprisingly, we conclude just the opposite for the gravity model
extended by a C-S LV term: the only tachyon- and ghost-free model
is the one with a time-like vector of the type $v^{a}=\left(\mu,\vec{0}\right)$.

In order to fulfill the task of analyzing the spectral properties
of these modified gravities, we follow the lines of previous works
\cite{our,extending the spin projection operators,Chern-Simons gravity}
and build up an orthonormal basis of operators that splits the fundamental
fields into their individual degrees of freedom. With this basis,
the attainment of the propagator is feasible and it is possible to
identify the excitation modes of the model. Also, we may suitably
interpret them as unitary representation of the subgroup that survives
from the Lorentz breakdown and, in this manner, as propagating particles.
Furthermore, we obtain conditions on the parameters of the Lagrangian
so that we may ensure propagation of non-tachyonic and non-ghost modes.

This paper is organized as follows: In Sec. \ref{sec:Lorentz-Breaking-Gravity},
we introduce our notations, conventions, and we start off with a general
Lagrangian including the Einstein-Hilbert term, (curvature)$^{2}$-terms
and the LV extensions given by the C-S and R-C terms. We pursue the
attainment of the propagators in Sec. \ref{sec:Attainment-of-the}.
The particle spectrum of the model is obtained in Sec. \ref{sec:Spectral-Consistency-Analysis},
where we also discuss the tachyon- and ghost-free conditions. In Sec.
\ref{sec:Concluding-Remarks}, we present our Concluding Remarks.
The operator basis, suitable for the attainment of the propagators,
is presented in the Appendix \ref{sec:Degree-of-Freedom-Projection-Operators}.
In the Appendix \ref{sec:Alternative-Method-for}, we present an alternative
method to obtain the propagator so that we can confirm the results
found out in Sec. \ref{sec:Attainment-of-the}.

\section{Higher Derivative Gravity Modified by LV Terms \label{sec:Lorentz-Breaking-Gravity}}

Let us start off our analysis by considering the Einstein-Hilbert
Lagrangian modified by higher derivative terms and C-S and R-C Lorentz-violating
actions,

\begin{equation}
\mathcal{L}=\sqrt{-g}\left(\alpha R+\beta R_{ab}R^{ab}+\gamma R^{2}\right)+\mu\mathcal{L}_{CS}+\lambda\mathcal{L}_{RC},\label{eq:lagrangian}\end{equation}
where $\alpha$, $\beta$, $\gamma$, $\mu$ and $\lambda$ are arbitrary
parameters. The Chern-Simons term is given by

\begin{equation}
\mathcal{L}_{CS}=-\frac{1}{2}\epsilon^{abcd}v_{a}\Gamma_{cf}^{e}\left(\partial_{b}\Gamma_{ed}^{f}+\frac{2}{3}\Gamma_{bg}^{f}\Gamma_{de}^{g}\right),\end{equation}
whereas the Ricci-Cotton Lagrangian reads\begin{equation}
\mathcal{L}_{RC}=\varepsilon^{abcd}v_{d}R_{ae}D_{b}R_{c}^{\ e}.\end{equation}

The background vector $v^{a}$ is an embedding coordinate which is
assumed to transform as a four-vector under observer (or passive)
Lorentz transformations; however it is invariant under particle (or
active) transformations. Consequently, observer Lorentz symmetry is
preserved, although particle Lorentz symmetry is broken (for further
discussion see \cite{ColladayKost}). 

At this point, it is worth remarking that this LV mechanism is rather
different from another recent proposal for a LV gravity model: the
so-called Horava-Lifshitz gravity \cite{Horava,Sotiriou2009}. Although
both of them fall into a class of non-Lorentz invariant models, in
the Horava-Lifshitz gravity the breaking of Lorentz invariance is
implemented by endowing space-time with a preferred foliation of three-dimensional
spacelike surfaces. This defines the splitting of coordinates into
space and time and explicitly breaks general covariance down to the
subgroup of coordinate transformations:\textbf{\begin{equation}
\mathbf{x}\mapsto\tilde{\mathbf{x}}\left(t,\mathbf{x}\right),\quad t\mapsto\tilde{t}\left(t\right).\end{equation}
}The LV induced by a background vector is expected to be of very small
magnitude and the effects of the LV terms are tightly constrained
by several experiments \cite{Liberati2009}. The Horava-Lifshitz gravity,
on the other hand, is expected to flow to a relativistic regime at
low-energies.

Throughout the paper, we shall use Latin letters ($a,b,c,\ldots$)
for space-time index and adopt the plus-minus convention for the Minkowski
metric $\eta_{ab}=\mbox{diag}\left(1,-1-1-1\right).$ Also, we shall
follow the conventions: $R_{\ bcd}^{a}=\partial_{c}\Gamma_{bd}^{a}+\Gamma_{ce}^{a}\Gamma_{bd}^{e}-\left(c\leftrightarrow d\right)$,
$\Gamma_{bc}^{a}=\frac{1}{2}g^{ad}\left(\partial_{b}g_{dc}+\partial_{c}g_{db}-\partial_{d}g_{bc}\right)$,
$R_{bd}=R_{\ bad}^{a}$, $R=g^{bd}R_{bd}$. We stress that we are
working with the second order formalism and possible torsion effects
are not included in our study.

The R-C term is, in the linearized limit, a sort of a higher derivative
of the C-S term, as it shall soon become evident. Some properties
of the electromagnetic counterpart of this R-C term in 3D are discussed
in \cite{DeserJackiw}, where it is argued that a higher derivative
term no longer preserves the topological properties of the C-S term.
It is not known to the authors of this paper an extensive study of
the R-C term in 3D gravity. In principle, one could choose different
background vectors for the C-S and R-C terms. We shall, however, adopt
a minimalist point of view, where the LV arises from just one background
vector, $v_{a}$. A conclusive answer to this issue can only be settled
with a more fundamental approach where the mechanism for spontaneous
symmetry breaking is clearly defined. 

By means of the weak-field approximation, $g_{ab}=\eta_{ab}+h_{ab}$,
we are able to write the quadratic Lagrangian, up to total derivatives,
as

\begin{eqnarray}
\mathcal{L}_{\left(2\right)} & = & \frac{\alpha}{2}\left(-\frac{1}{2}h^{ab}\square h_{ab}+\frac{1}{2}h\square h-h\partial_{a}\partial_{b}h^{ab}+h^{ab}\partial_{a}\partial_{c}h_{\ b}^{c}\right)\label{eq:LinearizedLagrangian}\\
 & + & \frac{\beta}{4}\left(h^{ab}\square^{2}h_{ab}+h\square^{2}h-2h\square\partial_{a}\partial_{b}h^{ab}-2h^{ab}\square\partial_{a}\partial_{c}h_{b}^{\ c}+2h^{ab}\partial_{a}\partial_{b}\partial_{c}\partial_{d}h^{cd}\right)\nonumber \\
 & + & \gamma\left(h\square^{2}h-2h\square\partial_{a}\partial_{b}h^{ab}+h^{ab}\partial_{a}\partial_{b}\partial_{c}\partial_{d}h^{cd}\right)\nonumber \\
 & + & \frac{\mu}{4}\epsilon^{abcd}v_{a}h_{d}^{\ e}\partial_{b}\left(\partial_{e}\partial_{f}h_{\ c}^{f}-\square h_{ec}\right)\nonumber \\
 & + & \frac{\lambda}{4}\epsilon^{abcd}v_{a}h_{d}^{\ e}\square\partial_{b}\left(\partial_{e}\partial_{f}h_{\ c}^{f}-\square h_{ec}\right).\nonumber \end{eqnarray}

For a non-trivial $v^{\mu}$, the non-linearized model \eqref{eq:lagrangian}
is not invariant under active diffeomorphism transformations. However,
the authors of \cite{JackiwPi} show that diffeomorphism symmetry
is recovered dynamically by means of the equations of motion. This
fact is reflected in a gauge symmetry of the action. The model is
invariant under the field transformation:\begin{equation}
h_{ab}'=h_{ab}+\partial_{a}\xi_{b}+\partial_{b}\xi_{a}.\label{eq:gauge transformation}\end{equation}
Other fundamental symmetry lost is CPT-invariance due to the fact
that we have a LV tensor (vector) with an odd number (namely one)
of indices \cite{Kost2004}.

One readily realizes that, besides derivatives and Minkowski metric,
there is the emergence of the Levi-Civita tensor and the LV background
vector $v_{a}$ in the wave operator \eqref{eq:LinearizedLagrangian},
which cannot be accommodated into the well-known Barnes-Rivers operators
\cite{rivers1964}. An analogous situation occurs in parity-breaking
theories in (1+2)-D, where there is the need of an extension of the
spin-operators basis in order to handle the Levi-Civita tensor \cite{extending the spin projection operators}.
Such an extension is also necessary in the present problem, and this
is the content of the Appendix \ref{sec:Degree-of-Freedom-Projection-Operators}.

\section{Writing Down the Propagators\label{sec:Attainment-of-the}}

A first step for the attainment of the propagator is to cast (up to
total derivatives) the linearized Lagrangian \eqref{eq:LinearizedLagrangian}
under the form\begin{equation}
\mathcal{L}_{\left(2\right)}=\frac{1}{2}h^{ab}\mathcal{O}_{ab,cd}h^{cd},\end{equation}
where the wave-operator, $\mathcal{O}_{ab,cd}$, in momentum space
reads as below:

\begin{eqnarray}
\mathcal{O}_{ab,cd} & = & \frac{1}{2}p^{2}\left(\alpha+\beta p^{2}\right)P\left(2\right)_{ab,cd}+\frac{1}{4}\left(\mu p^{2}+\lambda p^{4}\right)S_{ab,cd}+p^{2}\left[\left(6\beta+2\gamma\right)p^{2}-\alpha\right]P_{11}\left(0\right)_{ab,cd}.\label{eq:waveOperator}\end{eqnarray}

With the degree-of-freedom basis of operators discussed in the Appendix
\ref{sec:Degree-of-Freedom-Projection-Operators}, the wave operator
can be expanded as

\begin{equation}
\mathcal{O}_{ab,cd}=\sum_{J,ij}a_{ij}\left(J\right)P_{ij}\left(J\right)_{ab,cd}.\label{eq:waveOperatorExpansion}\end{equation}
Let us clarify the notation. The $a_{ij}\left(J\right)$ are the coefficients
in the wave operator expansion. The diagonal operators, $P_{ii}\left(J\right)$,
are projectors for each of the degrees of freedom of the spin $\left(J\right)$-sectors
of the field $h_{ab}$, while the $P_{ij}(J)$, with $i\neq j$, are
mappings between the projectors $P_{ii}\left(J\right)$ and $P_{jj}\left(J\right)$.
The attribution of projectors and mapping operators comes from the
orthonormal relation that these operators satisfy:

\begin{equation}
\sum_{cd}P_{ij}\left(I\right)_{ab,cd}P_{kl}\left(J\right)_{\ \ ,ef}^{cd}=\delta_{jk}\delta^{IJ}P_{il}\left(I\right)_{ab,ef}.\end{equation}
The property of decomposion of unity is expressed by: \begin{equation}
\sum_{i,J}P_{ii}\left(J\right)_{ab,cd}=\delta_{ab,cd}.\end{equation}

The coefficients $a_{ij}\left(J\right)$ can be arranged as matrices
that represent the contribution to the particular spin $\left(J\right)$.
When these matrices are non-singular, the saturated propagator is
given by:

\begin{equation}
\Pi=i\sum_{J,i,j}a_{ij}^{-1}\left(J\right)\mathcal{J}^{*ab}P_{ij}\left(J\right)_{ab,cd}\mathcal{J}^{cd},\label{eq:propagator}\end{equation}
where $\mathcal{J}^{ab}$ are physical sources that couple to the
propagator under consideration.

However, Lagrangian \eqref{eq:lagrangian} in its linearized form
is invariant under some local transformations of the fields \eqref{eq:gauge transformation}.
Gauge invariance imply that the coefficient matrices become degenerate.
In Ref. \cite{neville}, it is shown that the correct gauge invariant
propagator is obtained by taking the inverse of any largest non-degenerate
sub-matrix, denoted by $A_{ij}\left(J\right)$, which is next saturated
with sources.

For the model \eqref{eq:lagrangian}, the coefficients $a_{ij}\left(J\right)$
form the $\left(2\times2\right)$ spin-$0$, $\left(3\times3\right)$
spin-$1$ and $\left(5\times5\right)$ spin-$2$ matrices corresponding
to the 10 degrees of freedom contained in the metric field: 

\begin{equation}
a(0)=\left(\begin{array}{cc}
2\left(3\beta+\gamma\right)p^{4}-\alpha p^{2} & 0\\
0 & 0\end{array}\right),\end{equation}

\begin{eqnarray}
a\left(2\right) & = & \frac{p^{2}}{2}\left(\begin{array}{ccccc}
\beta p^{2}+\alpha & 0 & 0 & -i\sqrt{p_{*}^{2}}\left(\mu+\lambda p^{2}\right) & 0\\
0 & \beta p^{2}+\alpha & -i\frac{\sqrt{p_{*}^{2}}}{2}\left(\mu+\lambda p^{2}\right) & 0 & 0\\
0 & i\frac{\sqrt{p_{*}^{2}}}{2}\left(\mu+\lambda p^{2}\right) & \beta p^{2}+\alpha & 0 & 0\\
i\sqrt{p_{*}^{2}}\left(\mu+\lambda p^{2}\right) & 0 & 0 & \beta p^{2}+\alpha & 0\\
0 & 0 & 0 & 0 & \beta p^{2}+\alpha\end{array}\right).\label{eq:spin 2 matrix}\end{eqnarray}

The spin-$0$ matrix is degenerate, while the spin-$1$ matrix vanishes
identically, manifesting the gauge symmetry of the model. For the
spin-$2$, we have sort of a block-diagonal structure of the (1-4)-sector,
(2-3)-sector and the $5$-sector. This allows an inversion of each
sector separately. Their inverses, needed for the attainment of the
propagators, with the degeneracies duly extracted, are given by

\begin{equation}
A\left(0\right)=\frac{1}{p^{2}\left(2\left(3\beta+\gamma\right)p^{2}-\alpha\right)},\label{eq:spin0}\end{equation}

\begin{equation}
a_{14}^{-1}\left(2\right)=\frac{2}{p^{2}\Delta_{14}}\left(\begin{array}{cc}
\beta p^{2}+\alpha & i\sqrt{p_{*}^{2}}\left(\mu+\lambda p^{2}\right)\\
-i\sqrt{p_{*}^{2}}\left(\mu+\lambda p^{2}\right) & \beta p^{2}+\alpha\end{array}\right),\label{eq:spin2-14}\end{equation}

\begin{equation}
a_{23}^{-1}\left(2\right)=\frac{2}{p^{2}\Delta_{23}}\left(\begin{array}{cc}
\beta p^{2}+\alpha & i\frac{\sqrt{p_{*}^{2}}}{2}\left(\mu+\lambda p^{2}\right)\\
-i\frac{\sqrt{p_{*}^{2}}}{2}\left(\mu+\lambda p^{2}\right) & \beta p^{2}+\alpha\end{array}\right),\label{eq:spin2-23}\end{equation}

\begin{equation}
a_{5}^{-1}\left(2\right)=\frac{2}{p^{2}\left(\alpha+\beta p^{2}\right)},\label{eq:spin2-5}\end{equation}
where $\Delta_{14}=\left(\beta p^{2}+\alpha\right)^{2}-p_{*}^{2}\left(\mu+\lambda p^{2}\right)^{2}$,
$\Delta_{23}=\left(\beta p^{2}+\alpha\right)^{2}-\frac{p_{*}^{2}}{4}\left(\mu+\lambda p^{2}\right)^{2}.$
An alternative method to obtain the propagator is given in the Appendix
\ref{sec:Alternative-Method-for}. This supplementary method is somewhat
algebraically simpler, so that it may be used as a check of our calculations;
unfortunately, it provides less physical insights on the properties
of the propagating modes.

At a first glance, one observes that the spin-$0$ sector is unaffected
by the LV term. An explanation for this fact comes from the observation
that the LV operator, $S_{ab,cd}$, lives entirely in the spin-$2$
sector, as pointed out in the Appendix \ref{sec:Degree-of-Freedom-Projection-Operators}. 

However, the spin-$2$ sector is completely split for this LV theory.
In relativistic field theory, the spin of the particles is defined
in connection with the unitary representations of the little group
defined by a representative momentum of the class which the particle
belongs to. A background vector that breaks the isotropy of space-time
may therefore modify the spin structure of the theory. This is seen
explicitly in the spin-$2$ sector, where some degrees of freedom
may propagate independently.

An analogous phenomenon occurs in (1+2)-D in gravity theories with
parity-breaking terms. Since spin is represented by a pseudo-scalar
operator in 3-D, there must be a doublet of spins with the same absolute
value for the mass, $\left|m\right|$, so that an irreducible representation
of the Lorentz group extended by time-inversion and parity transformations
be set up. One the other hand, in a parity-breaking theory, such as
gravity theories added to a Chern-Simons parity-breaking term, this
doublet structure is lost and each spin component acquires a different
mass, which we interpret as degrees of freedom propagating independently.
For more details, see \cite{Chern-Simons gravity}.

The close link between the four-dimensional LV theory and the three-dimensional
parity-breaking theory is established when one considers the LV vector
with spatial components $v^{\mu}=\left(0,\vec{v}\right)$. In this
case, there is a symmetry breaking\begin{equation}
SO\left(1,3\right)\rightarrow SO\left(1,2\right);\end{equation}
therefore, the massive particles should not be any longer defined
by the embedding of $SU\left(2\right)\hookrightarrow SO\left(1,3\right)$,
but rather by $U\left(1\right)\hookrightarrow SO\left(1,2\right)$.
Furthermore, if the symmetry of the model considered is extended by
parity conservation, the particles can be defined as doublets of spins
$\left(s,-s\right)$, since, by parity operation, the spin $s$ is
mapped onto $-s$. This is the reason why the 5-dimensional matrix
\eqref{eq:spin 2 matrix} is split into the direct sum of the $U(1)$
fundamental representations and their complex conjugate (or equivalently
by $SO\left(2\right)$ representations). In this case, the new emerging
degrees of freedom can be enumerated by the usual $U\left(1\right)$
multiplication rule coming from group theory:\begin{equation}
\left(1\oplus-1\oplus0\right)\otimes\left(1\oplus-1\oplus0\right)=\left(3\times\underline{0}\oplus2\times\underline{1}\oplus2\times-\underline{1}\oplus\underline{2}\oplus-\underline{2}\right).\end{equation}

The three blocks appearing in the spin-$2$ sector may be viewed as
coming from a sort of dimensional reduction. If it is assigned the
$0$-helicity for the $\tau$-operator, the spin-$2$ operator decomposition
\eqref{eq:P(2)11}-\eqref{eq:P(2)55} can be compared with the three-dimensional
operators defined in \cite{extending the spin projection operators}.
For example: the $P_{11}\left(2\right)$ and $P_{44}\left(2\right)$
are clearly $P\left(2^{++}\right)$ and $P\left(2^{--}\right)$ of
the paper \cite{extending the spin projection operators}, respectively.
In this case, the spin-$2$ is broken to a {}``three-dimensional''
spin-$2$, which corresponds to the (1-4)-sector ($P\left(2\right)=P_{11}\left(2\right)+P_{44}\left(2\right)$);
spin-$1$, which corresponds to the (2-3)-sector ($P\left(1\right)=P_{22}\left(2\right)+P_{33}\left(2\right)$)
and spin-$0$, which corresponds to the $5$-sector ($P\left(0\right)=P_{55}\left(2\right)$).

\section{Analysis of the Spectral Consistency\label{sec:Spectral-Consistency-Analysis}}

In this Section, we analyze the spectral consistency of the model.
As a result of our study, we shall be able to impose conditions on
the parameters of Lagrangian \eqref{eq:lagrangian} in order to inhibit
the propagation of unphysical modes, that is, ghosts and tachyons.

In a quantum field theory with relativistic invariance, it is known
that the condition for absence of tachyon is $m^{2}\geq0$, where
$p^{2}=m^{2}$ appears as a pole of a given propagator. Also, the
statement for absence of ghosts reads

\begin{equation}
\Im\text{Res}(\Pi|_{p^{2}=m^{2}})>0.\label{eq:ghost_tachyon conditions for massive poles}\end{equation}
In the projection operators formalism, we can take advantage from
the general decomposition of the spin projection operator,\begin{equation}
P_{ij}\left(J\right)=\left(-1\right)^{P}\psi^{\left(i\right)}\psi^{\left(j\right)},\end{equation}
where $P$ is the parity related to the spin operator, to rewrite
the propagator \eqref{eq:propagator} as\begin{eqnarray}
\Pi & = & i\left(-1\right)^{P}\sum_{J,i,j,m^{2}}J_{i}^{*}A_{ij}\left(J,m^{2}\right)J_{j}\left(p^{2}-m^{2}\right)^{-1},\end{eqnarray}
where $J_{j}=\psi_{cd}^{\left(j\right)}J^{cd}$ and $A_{ij}\left(J,m^{2}\right)$
are the inverse sub-matrices with the pole extracted. Therefore, the
positiveness condition \eqref{eq:ghost_tachyon conditions for massive poles},
for arbitrary sources, is ensured by the positiveness of the eigenvalues
of the $A_{ij}\left(J,m^{2}\right)$ matrix.

Nevertheless, it can be shown that these matrices, for massive poles,
have only one non-vanishing eigenvalue at the pole, which is equal
to the trace of $A(J,m^{2})$. Therefore, the condition for absence
of ghosts for each spin is reduced to:

\begin{equation}
(-1)^{P}\mbox{tr}A(J,m^{2})|_{p^{2}=m^{2}}>0.\label{eq:ghost condition in terms of the trace}\end{equation}

On the other hand, in a LV theory, these conditions must be carefully
reassessed, since, for example, the dispersion relation is modified
and can be more general than simply $p^{2}=m^{2}$. In this case,
it is interesting to characterize tachyon and ghost excitations even
if Lorentz symmetry is no longer present. Due to subtleties that appear
for different choices of the background vector, we postpone the discussion
of the characterization of tachyon and ghost excitation modes in each
considered case.

The spin-$0$ sector is unaffected by the LV terms. So, the usual
constraints remain in order. The spin-$0$ massive pole keeps unchanged
the expression $m^{2}=\frac{\alpha}{6\beta+2\gamma}$, and the ghost-
and tachyon-free conditions are given by\begin{eqnarray}
\mbox{Spin-}\mathbf{0}: &  & \alpha>0;\ 3\beta+\gamma>0.\end{eqnarray}

Interestingly, the $5$-sector of the spin-$2$ shows up the massive
graviton yielded by the $R_{ab}R^{ab}$ term. However, the massive
pole, $m^{2}=-\frac{\alpha}{\beta}$, results in contradictory conditions
with positiveness of the Newton's constant, if one requires ghost
and tachyon absence:\begin{eqnarray}
\mbox{Sector-}5:\  & \alpha<0;\ \beta>0.\end{eqnarray}

Being inconsistent with unitarity requirements, we shall henceforth
assume that $\beta=0$, that is, the absence of the Ricci squared
term. Before analyzing the general tachyon- and ghost-free conditions
for the general case of the theory with the R-C and the C-S term,
we think it is instructive to study the simplest case, where the Ricci-Cotton
term is switched off ($\lambda=0$). In this situation, the matrix
corresponding to the 1-4-sector is cast as\begin{equation}
a_{14}^{-1}\left(2\right)=\frac{2}{\mu^{2}p^{2}\left(\frac{\alpha^{2}}{\mu^{2}}-p_{*}^{2}\right)}\left(\begin{array}{cc}
\alpha & i\mu\sqrt{p_{*}^{2}}\\
-i\mu\sqrt{p_{*}^{2}} & \alpha\end{array}\right).\label{eq:a_14-EH+CS}\end{equation}

The particle content of the model is set up by the poles of the propagator.
For the 1-4-sector of the spin-$2$ matrix, the poles are given by
the roots of\begin{equation}
\frac{\alpha^{2}}{\mu^{2}}-\left(\left(v\cdot p\right)^{2}-v^{2}p^{2}\right)=0.\end{equation}
So, the solution to this equation for the energies of the modes reads:\begin{eqnarray}
 &  & p_{0}=\frac{\left|\vec{p}\right|}{\left|\vec{v}\right|}\left[v_{0}\cos\theta\pm\sqrt{\frac{\alpha^{2}}{\mu^{2}\vec{p}^{2}}-v^{2}\sin^{2}\theta}\right],\label{eq:relacao de dispersao1-4}\end{eqnarray}
where $\theta$ is the angle between $\vec{v}$ and $\vec{p}$. One
should notice that it is necessary that $\left|\vec{v}\right|\neq0$
for the presence of massive poles. At this stage, it is interesting
to open up the discussion of the interpretation of the modes with
negative energy, so as to ensure that no tachyons are present.

In field theories with Lorentz invariance, the dispersion relations
read like $p_{0}=\pm E\left(\vec{p}\right)$, with $E\left(\vec{p}\right)=\sqrt{\vec{p}^{2}+m^{2}}$.
The negative energy solutions are naturally incorporated into the
quantum description with the definition of the causal Feynman propagator
that takes into account the contribution of a negative energy mode
as being the annihilation of a positive energy mode. Also, the CPT-invariance
of the model suggests the interpretation of a negative energy mode
propagating to the past as being a positive energy mode propagating
to the future.

However, at the present case, CPT-invariance is lost and the dispersion
relations can be in general of the form $p_{0}=E_{\pm}$ with $\left|E_{+}\right|\neq\left|E_{-}\right|$
(e.g. \eqref{eq:relacao de dispersao1-4}). These different masses
for the particle and antiparticle can be problematic to the locality
of the quantum theory, as discussed in \cite{Greenberg}.

By analysing \eqref{eq:relacao de dispersao1-4}, we realize that,
in the case $v_{0}\neq0$, there are different masses for the particle
and anti-particle and it happens that the positivity cannot be ensured
to arbitrary directions of propagation, by virtue of the explicit
dependence on $\cos\theta$. We therefore conclude that it is incompatible
to assume $v_{0}\neq0$ and $\left|\vec{v}\right|\neq0$ simultaneously,
if we wish to ensure a consistent quantum theory.

In the particular case $v_{0}=0$, the poles are given by:

\begin{equation}
p_{0}=\pm\frac{1}{\left|\vec{v}\right|}\sqrt{\frac{\alpha^{2}}{\mu^{2}}+\vec{v}^{2}\vec{p}^{2}\sin^{2}\theta}.\end{equation}
In this case, the dispersion relation is like $p_{0}=\pm E$ and,
potentially, the interpretation of the negative energy modes could
be consistently associated to the anti-particles. In fact, in spite
of the lack of CPT-invariance of the model, as discussed in Sec. \ref{sec:Lorentz-Breaking-Gravity},
in the case of a background vector with spatial components, the breaking
of Lorentz symmetry leaves a residual $SO\left(1,2\right)$ symmetry.
This can also be seen by the strict relation between this massive
spin-$2$ mode and the massive graviton that appears in the topological
massive gravity (TMG) in 3D. This forces an interpretation of the
negative energy modes, that are consistently discussed in the three-dimensional
TMG. We know that, in the four-dimensional model, a $C\tilde{P}T$
symmetry remains, where $C$ and $T$ are the usual charge conjugation
and time reversal operation, whereas $\tilde{P}$ is a discrete improper
Lorentz transformation that reverses only one spatial component in
the orthogonal plane defined by the LV vector, $\vec{v}$. The $\tilde{P}$
is the usual parity operation in $\left(1+2\right)$-D. We therefore
conclude that the negative energy mode could be consistently interpreted
as a positive energy mode with opposite spin polarization. The fact
that this excitation is non-tachyonic is a result that is in agreement
with previous discussion in the literature \cite{GravitonExcitations}.

To conclude the spectral analysis, one must ensure the positivity
of the residues at the poles of the propagators. First of all, one
should observe that, off-shell, the residue matrix $a_{14}^{-1}\left(2\right)$
(eq. \eqref{eq:a_14-EH+CS}) has two distinct eigenvalues, namely
$C\left(\alpha+\mu\sqrt{p_{*}^{2}}\right)$ and $C\left(\alpha-\mu\sqrt{p_{*}^{2}}\right)$,
with $C=2\left[\mu^{2}p^{2}\left(\frac{\alpha^{2}}{\mu^{2}}-p_{*}^{2}\right)\right]^{-1}$.
On-shell, that is, at $p_{*}^{2}=\frac{\alpha^{2}}{\mu^{2}}$, only
one of these two eigenvalues survives depending on the sign of $\frac{\alpha}{\mu}$.
Physically, this corresponds to the fact that just one spin polarization
propagates: $+2$ for $\frac{\alpha}{\mu}>0$, or $-2$ for $\frac{\alpha}{\mu}<0$.
This is of great resemblance with topologically massive gravity in
3D, where there is a propagation of single massive mode of helicity
$\pm2$, depending on the sign of the Chern-Simons term \cite{deser-jackiw-templeton}.

As previously discussed, the condition for the positivity of the eigenvalues
at the pole is simplified to the positivity of the trace \eqref{eq:ghost condition in terms of the trace}.
In this case,\begin{equation}
\mbox{Res}\left.\mbox{tr}a_{14}^{-1}\left(2\right)\right|_{p_{0}^{2}=E^{2}}=-\frac{2\alpha}{\mu^{2}\vec{v}^{2}p^{2}},\label{eq:ghostMassivo}\end{equation}
where, at the pole,

\begin{equation}
p^{2}=\frac{\alpha^{2}}{\mu^{2}\vec{v}^{2}}-\vec{p}^{2}\cos^{2}\theta.\end{equation}
For low momenta, $\vec{p}^{2}<\frac{\alpha^{2}}{\mu^{2}\vec{v}^{2}}$,
the condition for the propagation of a non-ghost mode, expressed in
eq. \eqref{eq:ghostMassivo}, dictates: \begin{equation}
\alpha<0.\label{eq:alpha>0}\end{equation}
One could wonder if the model would be consistent with $\alpha>0$
for specific choices of $\vec{p}^{2}\cos^{2}\theta$. We understand
that an explicit dependence of the consistency of the model on the
direction of propagation is an odd situation and so we discard this
possibility. In fact, for the case where the dynamics is restricted
to the plane orthogonal to $\vec{v}$, $\cos\theta$ vanishes identically
and $p^{2}$ is positive definite. Again, the requirement of a negative
Newton's constant is in agreement with previous consideration of the
relation of this model with TMG.

In the more general case, with the Ricci-Cotton term present $\left(\lambda\neq0\right)$,
the massive poles of the propagator of the spin (1-4)-sector are given
by the roots of 

\begin{eqnarray}
\Delta_{14} & = & \alpha^{2}-p_{*}^{2}\left(\mu+\lambda p^{2}\right)^{2}.\label{eq:denominadorRC}\end{eqnarray}
The condition for absence of ghost reads\begin{equation}
\mbox{Res }\left.\left(\mbox{tr}a_{14}^{-1}\left(2\right)\right)\right|_{p_{0}^{2}=E^{2}}=\frac{2\alpha\left(p_{0}^{2}-E^{2}\right)}{p^{2}\Delta_{14}}>0.\label{eq:ghostCondRC}\end{equation}
However, condition \eqref{eq:ghostCondRC} is doomed to propagate
ghost modes. The reason is simple: the denominator \eqref{eq:denominadorRC}
brings three massive poles, $\Delta_{14}=C\left(p_{0}^{2}-E_{1}^{2}\right)\left(p_{0}^{2}-E_{2}^{2}\right)\left(p_{0}^{2}-E_{3}^{2}\right)$,
with $E_{1}^{2}<E_{2}^{2}<E_{3}^{2}$ or two massive poles (depending
on the choice of $v^{\mu}$), $\Delta_{14}=C'\left(p_{0}^{2}-E_{1}^{2}\right)\left(p_{0}^{2}-E_{2}^{2}\right)$,
with $E_{1}^{2}<E_{2}^{2}$. In either cases, if $\mbox{tr}\left.a_{14}^{-1}\left(2\right)\right|_{p_{0}^{2}=E_{1}^{2}}$
is positive (negative), then $\mbox{tr}\left.a_{14}^{-1}\left(2\right)\right|_{p_{0}^{2}=E_{2}^{2}}$
is negative (positive). So, the positivity of the residue in all poles
cannot be ensured simultaneously. Such a phenomenon is ubiquitous
in theories with higher derivatives, when more than one massive pole
in the same spin sector usually brings ghosts. In view of this problem,
we are bound to take $\lambda=0$, eliminating then the R-C term.

The spectral consistency analysis for the (2-3)-sector follows the
same reasoning of the (1-4)-sector, and the main results remains unchanged.
This finishes the discussion of quantum consistency for the massive
poles. Let us tackle now the spectrum consistency analysis for the
massless pole.

\subsection{The Massless Graviton Pole\label{sub:The-Massless-Graviton} }

The massless poles must be handled with extra care. At a first sight,
one sees that the basis of operators is ill-defined for light-like
momenta. However, one can show that the physical sources satisfy constraints
($p_{a}\mathcal{J}^{ab}=0$) due to gauge symmetries of the model.
These constraints make the saturated propagator a well-defined expression
even for light-like momenta. In fact, the surviving structures when
the projectors are saturated by the sources at light-like momenta
($p^{2}=0$) are given by:

\begin{subequations}\begin{eqnarray}
P_{11}\left(0\right)_{ab,cd} & = & \frac{1}{2}\eta_{ac}\eta_{cd}+\mbox{t.d.n.c.res.},\\
P_{11}\left(2\right)_{ab,cd} & = & \frac{1}{2}\left(\rho_{ac}\sigma_{bd}+\rho_{ad}\sigma_{bc}+\sigma_{ac}\rho_{bd}+\sigma_{ad}\rho_{bc}\right)+\mbox{t.d.n.c.res.},\\
P_{22}\left(2\right)_{ab,cd} & = & \mbox{t.d.n.c.res.},\\
P_{33}\left(2\right)_{ab,cd} & = & \mbox{t.d.n.c.res.},\\
P_{44}\left(2\right)_{ab,cd} & = & \frac{1}{2}\left(\rho_{ab}\rho_{cd}+\sigma_{ab}\sigma_{cd}\right)-\frac{1}{2}\left(\rho_{ab}\sigma_{cd}+\sigma_{ab}\rho_{cd}\right)+\mbox{t.d.n.c.res.},\\
P_{55}\left(2\right)_{ab,cd} & = & \frac{1}{2}\eta_{ac}\eta_{cd}+\mbox{t.d.n.c.res.},\end{eqnarray}
\end{subequations}where t.d.n.c.res. is an acronym for {}``terms
that do not contribute to the residue''. From the expression of $\tau_{ab}$
in terms of $v_{a}$ and $p_{a}$ (equation \eqref{eq:tau}), one
can show that $\tau_{ab}\mathcal{J}^{bc}=0$. This is why $P_{22}\left(2\right)$
and $P_{33}\left(2\right)$ do not contribute to the residues, and
only the $\rho$'s $\sigma$'s and $\eta$'s appear in the final expression.

Another remark that should be made is that, for non-vanishing mapping
operators, the projection contribution can be diagonalized $\left(\tilde{P}_{ij}=A_{ik}P_{kl}A_{lj}^{T}\right)$
in such a way that the rotated projectors contribute with the corresponding
eigenvalue. In this way, the propagators can be written as

\begin{eqnarray}
\Im\text{Res}(\Pi|_{p^{2}=0}) & = & \mathcal{\tilde{J}}^{*ab}\left[\frac{2}{\alpha-\mu\left|v\cdot p\right|}\tilde{P}_{11}\left(2\right)_{ab,cd}+\frac{2}{\alpha+\mu\left|v\cdot p\right|}\tilde{P}_{44}\left(2\right)_{ab,cd}\right]\mathcal{\tilde{J}}^{cd}.\label{eq:imResp2_0}\end{eqnarray}
It is interesting to notice that the contribution of the mapping operators
is responsible for breaking up the two degrees of freedom of the usual
graviton. In the case where the LV parameter, $\mu$, vanishes, we
recover the well-known graviton propagator in 4D. 

The origin of the splitting in the dynamics of these independent degrees
of freedom of the massless particle is analogous to the phenomenon
that occurs for massive particles in 3D, when there is a parity-breaking
term. There is indeed a great resemblance between massive particles
in 3D and massless particles in 4D, since they essentially share the
same representation structure of the Poincaré group. As in 3D, parity-breaking
terms yield different masses for particles that would propagate as
doublet of spins, as discussed in \cite{Chern-Simons gravity}. This
same role of the parity-breaking shows up in the massless graviton
propagator \eqref{eq:imResp2_0}, where the helicities modes are no
more related by CPT transformations.

To analyze unitarity, one should recall that the sources are arbitrary
and $\tilde{P}_{11}\left(2\right)$ and $\tilde{P}_{44}\left(2\right)$
are independent operators. In order to ensure the positivity of the
expression \eqref{eq:imResp2_0}, we must impose that the eigenvalues
are positive-definite, independently. This implies that\begin{equation}
\left|\mu\left(v\cdot p\right)\right|<\alpha.\label{eq:condicaoUnitariedadeMassless}\end{equation}

Condition \eqref{eq:condicaoUnitariedadeMassless} means, in particular,
that the unitarity constraint of our LV theory is in accordance with
the usual requirement of the positiveness of the Newton's constant
in 4D. However, such a condition is conflicting with the requirement
of \eqref{eq:alpha>0}. This is not a surprise for us in the light
of the close link between parity-breaking theories in 3D and LV theories
in 4D with a breaking vector with spatial components. In fact, it
is known that Chern-Simons massive gravities are consistent solely
if $\alpha<0$. In 3D, the massless graviton does not propagate and
there is no conflict with respect to unitarity. In a Chern-Simons
LV gravity in 4D, we conclude, therefore, that the propagation of
the massless graviton and the massive modes coming from the violation
of the Lorentz symmetry are not compatible if one enforces to respect
the requirement of propagating only non-ghost modes.

Though our efforts have been made to find the correct relations among
the free parameters and the LV background vector to ensure absence
of tachyons and ghosts, it must be emphasized that these are tree-level
conditions, valid at linear level, stem from our inspection of the
residues and poles of the free propagators. Interactions and loop
corrections might give rise to unphysical modes which, in turn, could
be suppressed by re-analyzing the spectrum and finding new conditions
among the loop-corrected parameters, so that ghosts and tachyons are
also eliminated from the loop-corrected action. Therefore, at non-linear
level, suppression of ghosts and tachyons has to be reassessed at
each order in perturbation theory.

\section{Concluding Remarks\label{sec:Concluding-Remarks}}

We have considered a general gravity Lagrangian with higher derivatives
and CPT/Lorentz-violating Chern-Simons and Ricci-Cotton term in the
second-order formalism. It was our interest to investigate the implication
of the CPT/Lorentz-symmetry violating terms in the unitary properties
of the model and, also, determine the particle spectrum of the theory.

With this aim, we have developed a basis of degree-of-freedom operators.
The proposed basis turns out to be algebraically convenient and still
exhibits a clear physical interpretation. The principle of defining
degree-of-freedom operators renders a great flexibility to identify
the physical content of the excitation modes. This generality is specially
interesting if the background vector is not fixed a priori, since,
in this case, there is no definition, in advance, for the particles.
In the case where Lorentz symmetry remains unbroken, these degree-of-freedom
operators will always rearrange to the usual spin projectors, as expected.
However, if there is a breaking of the Lorentz symmetry by a background
vector with only spatial components, $SO\left(1,2\right)$ symmetry
is residual. In this case, the degrees-of-freedom operators rearrange
in such a way that the planar-like excitations can be readily identified.
Or, loosely speaking, this represents a dynamical generation of the
spin-particle by using the degrees of freedom of the fields.

Extensions to include more general background tensors, as discussed
in the Standard Model Extension (SME), considered by Kostelecky and
collaborators \cite{Kost2004}, could be addressed too. In the case
of two linear independent background vectors, there will be a Lorentz
symmetry breaking, $SO\left(1,3\right)\rightarrow U\left(1\right)$,
and we expect that the excitations shall not be arranged in planar-like
modes. This situation may be relevant whenever LV is triggered by
a spontaneous supersymmetry breaking mechanism \cite{Belich,Belich2},
since another background vector may appears by means of a background
fermion condensation. Furthermore, we suppose that the degree-of-freedom
operators could be defined and the other Lorentz breaking vector should
be split into three orthogonal contributions: $p_{\mu}$, $e_{3\mu}$
and $e_{2\mu}$, according to the Gram-Schmidt decomposition. 

The breaking of the CPT-symmetry threatens the interpretation of the
negative energy modes that are common in quantum field theory. This
is the matter of a detailed discussion. We conclude that a background
vector, with $v_{0}\neq0$ and $\vec{v}\neq0$, brings up an inconsistent
quantum model, since the propagating modes does not have a positive-definite
energy for arbitrary directions of propagation. The situation where
$v_{0}=0$ and $\vec{v}\neq0$, revealed non-tachyonic excitations
after a revision of the role of the discrete symmetries. In spite
of this known fact, we showed, however, that these modes are negative-normed
states (ghosts) and with mass comparable to the Planck mass. These
fact warns for possible inconsistencies in the quantum version of
the model. Nevertheless, in the search for small deviations in low-energy
processes, where the problematic mode should not be excited, one may
obtain consistent results if the effective model is viewed as a relic
of a more fundamental (and consistent) theory. The last case, where
$v_{0}\neq0$ and $\vec{v}=0$, which was the subject of an investigation
in \cite{JackiwPi}, revealed, in the limiting case of Einstein-Hilbert
Chern-Simons gravity, no other propagating modes besides the massless
graviton. It is in agreement with the results by Jackiw and Pi that
the two polarizations of the graviton propagate independently. The
general behavior of (curvature)$^{2}$-terms is not altered by the
LV Chern-Simons term, in the sense that the $R^{2}$-term propagates
a nonghost massive spin-0 and the $\left(R_{ab}\right)^{2}$-term
propagates a massive spin-2 ghost. The higher derivative Ricci-Cotton
term brings unavoidable ghosts modes.

Another interesting pursuit is to tackle an analogous problem, but
in the first order formalism, where the vielbein and the spin connection
are understood to be independent fields. In such a situation, we expect
that the difficulty in the calculations shall be greatly increased,
since in the analogous problem with dynamical torsion there is a plethora
of propagating modes, $2^{+}$, $2^{-}$, $1^{+}$, $1^{-}$, $0^{+}$,
$0^{-}$ \cite{S-N,our} in which, with LV, the spin modes would be
split into spin polarization modes. In spite of this difficulty, this
problem would be rather interesting, since, in this formalism, there
is the possibility of considering, besides the usual Chern-Simons
term in the first order formalism $\epsilon^{abcd}v_{d}(R_{abef}\omega_{c}^{\ ef}+\frac{2}{3}\omega_{af}^{\ \ g}\omega_{bg}^{\ \ e}\omega_{ce}^{\ \ f})$,
other combinations that have not been considered previously: $\epsilon^{abcd}v_{a}T_{bc}^{\ \ e}R_{de}$,
$\epsilon^{abcd}v_{a}RT_{bcd}$ and $\epsilon^{abcd}v_{a}T_{be}^{\ \ e}R_{cd}$.
Also, torsion effects could reveal interesting consequences in this
scenario.
\begin{acknowledgments}
The authors are very grateful to Professors S. A. Dias and H. Belich
for the supporting discussions. Thanks are also due to Professor J.
A. Accioly for the encouragement in pursuing a deeper investigation
of our work. The authors B. P. D. and C. A. H. express their gratitude
to CNPq-Brazil and FAPERJ-Rio de Janeiro for the Graduate fellowships.
J. A. H. N. acknowledges CNPq for the invaluable financial support.
\end{acknowledgments}
\appendix

\section{Degree-of-Freedom Projection Operators\label{sec:Degree-of-Freedom-Projection-Operators}}

In order to analyze the particle spectrum of the model and the nature
of its associated particles, the attainment of the propagator becomes
a primary goal. With the propagator at hand, one reads off the masses
of the particles and may verify the positiveness of the residues of
the propagators at the correspond poles, so as to determine conditions
for the absence of ghosts. A suitable method to obtain the propagator
and identify the particle content of the models is the one based on
the spin projector operators \cite{rivers1964,neville,S-N,our}. In
four dimensions, in the second order formalism, one has the Barnes-Rivers
operators that form a complete orthonormal basis of operators for
models with Lorentz and CPT invariances:

\begin{subequations}

\begin{eqnarray}
P\left(2\right)_{ab,cd} & = & \frac{1}{2}\left(\theta_{ac}\theta_{bd}+\theta_{ad}\theta_{bc}\right)-\frac{1}{3}\theta_{ab}\theta_{cd};\\
P\left(1\right)_{ab,cd} & = & \frac{1}{2}\left(\theta_{ac}\omega_{bd}+\theta_{ad}\omega_{bc}+\theta_{bc}\omega_{ad}+\theta_{bd}\omega_{ac}\right);\\
P_{11}\left(0\right)_{ab,cd} & = & \frac{1}{3}\theta_{ab}\theta_{cd};\quad P_{12}\left(0\right)_{ab,cd}=\frac{1}{\sqrt{3}}\theta_{ab}\omega_{cd};\\
P_{21}\left(0\right)_{ab,cd} & = & \frac{1}{\sqrt{3}}\omega_{ab}\theta_{cd};\quad P_{22}\left(0\right)_{ab,cd}=\omega_{ab}\omega_{cd}.\end{eqnarray}

\end{subequations}

However, when one adds terms that spoil these symmetries, one must
enlarge the basis in order to accommodate the new operators that appear
in the linearized Lagrangian. Since we wish to keep the background
vector arbitrary, we define a complete basis of operators that splits
the fundamental fields according to their individual degrees of freedom.
This is more convenient because different choices for the background
vector may lead to different particle species, once the symmetry group
that survives from the breaking of Lorentz symmetry depends on the
explicit form of the background vector.

As a starting point to set up a suitable basis of degree-of-freedom
operators, we can decompose the Barnes-Rivers operators in order to
accommodate LV operators, such as

\begin{equation}
S_{ab,cd}=\frac{1}{2}\left(\theta_{ac}S_{bd}+\theta_{ad}S_{bc}+\theta_{bc}S_{ad}+\theta_{bd}S_{ac}\right),\label{eq:cs-operator}\end{equation}
where, $S_{ab}=\epsilon_{abcd}v^{c}\partial^{d}$, that comes from
the C-S and R-C LV terms (in the linearized version of the theory).

A first hint to solve this task is to notice that the $S_{ab,cd}$-operator
{}``lives'' entirely in the spin-$2$ sector. This is seen from
the fact that $S_{ab,cd}$ is annihilated by the spin-$1$ and spin-$0$
operators, but is remains unchanged by the spin-$2$ projection operator,
as shown in the following relations:

\begin{subequations}

\begin{eqnarray}
P\left(2\right)_{ab,cd}S_{\ \ ,ef}^{cd} & = & S_{ab,ef}\\
P\left(1\right)_{ab,cd}S_{\ \ ,ef}^{cd} & = & 0\\
P_{ij}\left(0\right)_{ab,cd}S_{\ \ ,ef}^{cd} & = & 0,\quad i,j=1,2.\end{eqnarray}

\end{subequations}

With these remarks, we shall pursue the task of the attainment of
the basis of degree-of-freedom operators for gravity models, but first
it is necessary to classify the building blocks.

\subsection{Building Blocks}

In a context of LV, there is the need of redefining the building blocks
of the projections operators. This can be motivated by the decomposition
of the Lorentz-breaking vector, $v^{a}$, into a term proportional
to the momentum and an orthogonal component,

\begin{equation}
v^{a}=\frac{\left(v\cdot p\right)}{p^{2}}p^{a}+\sqrt{\frac{p_{*}^{2}}{p^{2}}}e_{3}^{a},\label{eq:vetorV}\end{equation}
where $p^{a}$ is the relativistic four-momentum assumed to be time-like
and $p_{*}^{2}=\left(v\cdot p\right)^{2}-v^{2}p^{2}$. Nevertheless,
the whole discussion can take place in the case of a light-like momentum,
necessary for the analysis of the massless poles (Sec. \ref{sub:The-Massless-Graviton}).

The building blocks for the operators can be built up from $p_{a}$
and three other orthonormal space-like vectors. Without loss of generality,
we may choose one of them as $e_{3}^{a}=$$\sqrt{\frac{p^{2}}{p_{*}^{2}}}\left[v^{a}-\frac{\left(v\cdot p\right)}{p^{2}}p^{a}\right]$
and another two vectors, $e_{1}^{a}$ and $e_{2}^{a}$, orthogonal
to each other and orthogonal to $p^{a}$ and $e_{3}^{a}$. We assume
that $p_{*}^{2}\neq0$, since if $p_{*}^{2}=0$ implies that $v^{a}\parallel p^{a}$
and so the C-S and R-C terms necessarily vanish.

With these vectors, one may define the following projection operators:

\begin{subequations}

\begin{eqnarray}
\omega^{ab} & = & \frac{p^{a}p^{b}}{p^{2}},\label{eq:omega}\\
\rho^{ab} & = & -e_{1}^{a}e_{1}^{b},\\
\sigma^{ab} & = & -e_{2}^{a}e_{2}^{b},\\
\tau^{ab} & = & -e_{3}^{a}e_{3}^{b}=-\frac{1}{p_{*}^{2}}\left[p^{2}v^{a}v^{b}-\left(v\cdot p\right)\left(p^{a}v^{b}+v^{a}p^{b}\right)+\left(v\cdot p\right)^{2}\omega^{ab}\right].\label{eq:tau}\end{eqnarray}

\end{subequations}One should notice that the transverse operator,
$\theta^{ab}$, can be related to these operators by,

\begin{equation}
\theta^{ab}\equiv\eta^{ab}-\omega^{ab}=\rho^{ab}+\sigma^{ab}+\tau^{ab}.\label{eq:theta}\end{equation}

The operator for the LV Chern-Simons term, $S^{ab}=i\epsilon^{abcd}v_{c}p_{d}$,
may also be written in terms of the building blocks. Using \eqref{eq:vetorV}
and \eqref{eq:theta}, one actually shows that\begin{eqnarray}
S^{ab} & = & i\sqrt{p_{*}^{2}}\left(e_{1}^{a}e_{2}^{b}-e_{2}^{a}e_{1}^{b}\right).\label{eq:}\end{eqnarray}

With the basis \eqref{eq:omega}-\eqref{eq:tau} at hand, one could
carry out the task of getting the propagator of the Maxwell-Chern-Simons
LV theory, with eventually a massive term\begin{equation}
\mathcal{L}=-\frac{1}{4}F_{ab}F^{ab}-\frac{M^{2}}{2}A_{a}A^{a}+\frac{1}{4}\epsilon^{abcd}v_{a}A_{b}F_{cd}.\label{eq:MCS-Lagrangian}\end{equation}
This problem has been considered in \cite{Scarpelli-Belich,Adam-Klinkhamer}
using different approaches. Instead, let us move on to deeper waters
and tackle a related problem in gravity.

\subsection{Degree-of-freedom Operators for Gravity Models}

The spin-$2$ sector operators result from the decomposition of the
operator $P\left(2\right)_{ab,cd}=\frac{1}{2}\left(\theta_{ac}\theta_{bd}+\theta_{ad}\theta_{bc}\right)-\frac{1}{3}\theta_{ab}\theta_{cd}$.
A more efficient method to decompose this spin projection operator
into orthogonal sub-components can be achieved by its decomposition
in normalized eigenvectors, $\psi_{ab}^{\left(i\right)}$:\begin{equation}
P\left(2\right)_{ab,cd}=\sum_{i}\psi_{ab}^{\left(i\right)}\psi_{cd}^{\left(i\right)},\end{equation}
(for arbitrary spin, this decomposition must include a factor coming
from the parity $P$ of the spin sector: $P_{ij}\left(J\right)=\left(-1\right)^{P}\sum_{i}\psi^{\left(i\right)}\psi^{\left(i\right)}$),
where\begin{equation}
P\left(2\right)_{ab,cd}\psi^{\left(i\right)cd}=\psi_{ab}^{\left(i\right)}.\label{eq:eigvalueEquation}\end{equation}
With these eigenvectors at hand, one can define the full set of degree-of-freedom
projectors and mapping operators:\begin{equation}
P_{ij}=\psi^{\left(i\right)}\psi^{\left(j\right)}.\label{eq:projectors-eing}\end{equation}

A remark that has already been made in \cite{S-N} is that the projection-operator
basis is defined up to a rotation transformation. In fact, a rotation
of the eigenvector basis,\begin{equation}
\tilde{\psi}^{\left(i\right)}=\sum_{j}U_{ij}\psi^{\left(j\right)},\end{equation}
with $U_{ij}$ orthogonal ($UU^{T}=1$), redefine rotated projectors\begin{equation}
\tilde{P}_{ij}=\tilde{\psi}^{\left(i\right)}\tilde{\psi}^{\left(j\right)}=U_{ik}\psi^{\left(k\right)}U_{jl}\psi^{\left(l\right)}=U_{ik}P_{kl}U_{lj}^{T}.\end{equation}
But, the completeness relation may be used and, therefore, the spin-projectors
remain unaltered:\begin{equation}
\tilde{P}=\sum_{i}\tilde{\psi}^{\left(i\right)}\tilde{\psi}^{\left(i\right)}=\sum_{i,k,l}U_{ik}U_{il}\psi^{\left(k\right)}\psi^{\left(l\right)}=\sum_{i}\psi^{\left(i\right)}\psi^{\left(i\right)}=P.\end{equation}
 A convenient choice of eigenvectors helpful in the comparison with
the 3-dimensional analog theory is given below:

\begin{subequations}

\begin{eqnarray}
 & \psi_{ab}^{\left(1\right)} & =\frac{1}{\sqrt{2}}\left(e_{1a}e_{2b}+e_{2a}e_{1b}\right),\label{eq:psi1}\\
 & \psi_{ab}^{\left(2\right)} & =\frac{1}{\sqrt{2}}\left(e_{1a}e_{3b}+e_{3a}e_{1b}\right),\\
 & \psi_{ab}^{\left(3\right)} & =\frac{1}{\sqrt{2}}\left(e_{2a}e_{3b}+e_{3a}e_{2b}\right),\\
 & \psi_{ab}^{\left(4\right)} & =\frac{1}{\sqrt{2}}\left(\rho_{ab}-\sigma_{ab}\right),\\
 & \psi_{ab}^{\left(5\right)} & =\frac{1}{\sqrt{6}}\left(\rho_{ab}+\sigma_{ab}-2\tau_{ab}\right).\label{eq:psi5}\end{eqnarray}

\end{subequations}With the eigenvectors \eqref{eq:psi1}-\eqref{eq:psi5},
the degree-of-freedom operators are built up by the relation \eqref{eq:projectors-eing}.
The projection operators are then cast as: 

\begin{subequations}\begin{eqnarray}
P_{11}\left(2\right)_{ab,cd} & =\psi_{ab}^{\left(1\right)}\psi_{cd}^{\left(1\right)} & =\frac{1}{2}\left(\rho_{ac}\sigma_{bd}+\rho_{ad}\sigma_{bc}+\sigma_{ac}\rho_{bd}+\sigma_{ad}\rho_{bc}\right),\label{eq:P(2)11}\\
P_{22}\left(2\right)_{ab,cd} & =\psi_{ab}^{\left(2\right)}\psi_{cd}^{\left(2\right)} & =\frac{1}{2}\left(\rho_{ac}\tau_{bd}+\rho_{ad}\tau_{bc}+\tau_{ac}\rho_{bd}+\tau_{ad}\rho_{bc}\right),\\
P_{33}\left(2\right)_{ab,cd} & =\psi_{ab}^{\left(3\right)}\psi_{cd}^{\left(3\right)} & =\frac{1}{2}\left(\tau_{ac}\sigma_{bd}+\tau_{ad}\sigma_{bc}+\sigma_{ac}\tau_{bd}+\sigma_{ad}\tau_{bc}\right),\\
P_{44}\left(2\right)_{ab,cd} & =\psi_{ab}^{\left(4\right)}\psi_{cd}^{\left(4\right)} & =\frac{1}{2}\left(\rho_{ab}\rho_{cd}+\sigma_{ab}\sigma_{cd}\right)-\frac{1}{2}\left(\rho_{ab}\sigma_{cd}+\sigma_{ab}\rho_{cd}\right),\\
P_{55}\left(2\right)_{ab,cd} & =\psi_{ab}^{\left(5\right)}\psi_{cd}^{\left(5\right)} & =\frac{1}{6}\left[\rho_{ab}\rho_{cd}+\sigma_{ab}\sigma_{cd}+4\tau_{ab}\tau_{cd}+\rho_{ab}\sigma_{cd}+\sigma_{ab}\rho_{cd}\right.\label{eq:P(2)55}\\
 &  & \left.-2\left(\rho_{ab}\tau_{cd}+\tau_{ab}\rho_{cd}\right)-2\left(\sigma_{ab}\tau_{cd}+\tau_{ab}\sigma_{cd}\right)\right].\end{eqnarray}
\end{subequations}Accordingly, the mapping operators are given by
$P_{ij}\left(2\right)_{ab,cd}=\psi_{ab}^{\left(i\right)}\psi_{cd}^{\left(j\right)}$
$\left(i\neq j\right)$. For example,  \begin{equation}
P_{14}\left(2\right)_{ab,cd}=\psi_{ab}^{\left(1\right)}\psi_{cd}^{\left(4\right)}=\frac{1}{2}\left(e_{1a}e_{2b}+e_{2a}e_{1b}\right)\left(\rho_{cd}-\sigma_{cd}\right).\end{equation}
However, the mapping operators may be expressed in terms of $\epsilon$,
$v^{a}$, $\rho$ and $\sigma$ so as to facilitate the expansion
of the wave operator in terms of the degree-of-freedom operators \eqref{eq:waveOperatorExpansion}.
For $P_{14}\left(2\right)$, one can show that\begin{equation}
P_{14}\left(2\right)_{ab,cd}=\frac{1}{2}\epsilon^{efgh}\left(\rho_{ac}\sigma_{bf}\rho_{de}+\rho_{bd}\sigma_{af}\rho_{ce}-\sigma_{bd}\rho_{ae}\sigma_{cf}-\sigma_{ac}\rho_{be}\sigma_{df}\right)\frac{v_{g}p_{h}}{\sqrt{p_{*}^{2}}}.\end{equation}
The other mapping operators can be expressed, if desired, in an analogous
manner.

As already remarked, the linearized model is invariant under the gauge
transformation \eqref{eq:gauge transformation}. For this reason,
the spin-$1$ sector was fully suppressed from our discussion. In
spite of this, the spin-$1$ sector may be important for a gravitational
theory without gauge symmetry or with more propagating fields. For
completeness, we present in this appendix the spin-$1$ sector degree-of-freedom
operators. They are built up along the same lines as the spin-$2$
sector; the projectors are cast as:

\begin{subequations}\begin{eqnarray}
 & P_{11}\left(1\right)_{ab,cd} & =\frac{1}{2}\left(\rho_{ac}\omega_{bd}+\rho_{bc}\omega_{ad}+\rho_{ad}\omega_{bc}+\rho_{bd}\omega_{ac}\right),\\
 & P_{22}\left(1\right)_{ab,cd} & =\frac{1}{2}\left(\sigma_{ac}\omega_{bd}+\sigma_{bc}\omega_{ad}+\sigma_{ad}\omega_{bc}+\sigma_{bd}\omega_{ac}\right),\\
 & P_{33}\left(1\right)_{ab,cd} & =\frac{1}{2}\left(\tau_{ac}\omega_{bd}+\tau_{bc}\omega_{ad}+\tau_{ad}\omega_{bc}+\tau_{bd}\omega_{ac}\right).\end{eqnarray}
\end{subequations}

\section{An Alternative Method for the Attainment of Propagators\label{sec:Alternative-Method-for}}

In previous works, the derivation of the propagators for gravity models
with LV was also considered \cite{GravitonExcitations}, but using
the algebraic method without decomposition into degree-of-freedom
operators. Therefore, it is worthy to verify the consistency of our
resulting propagators. This comparison can be made by noting that
the C-S operator can be cast in the following form:

\begin{eqnarray*}
S_{ab,cd} & = & i\sqrt{p_{*}^{2}}\left[-2P_{14}\left(2\right)_{ab,cd}+2P_{41}\left(2\right)_{ab,cd}-P_{23}\left(2\right)_{ab,cd}+P_{32}\left(2\right)_{ab,cd}\right].\end{eqnarray*}

Such a decomposition clears up a fact, not obvious at a first sight,
that the operator $S_{ab,cd}$ is made up of two reducible components:

\begin{eqnarray}
\left(\tau S\right)_{ab,cd} & = & \frac{1}{2}\left(\tau_{ac}S_{bd}+\tau_{ad}S_{bc}+\tau_{bc}S_{ad}+\tau_{bd}S_{ac}\right)\label{eq:op1-tab-mult}\\
 & = & -i\sqrt{p_{*}^{2}}\left[P_{23}\left(2\right)_{ab,cd}-P_{32}\left(2\right)_{ab,cd}\right],\nonumber \\
\left(\theta\tau S\right)_{ab,cd} & = & \frac{1}{2}\left(\left(\theta_{ac}-\tau_{ac}\right)S_{bd}+\left(\theta_{ad}-\tau_{ad}\right)S_{bc}+\left(\theta_{bc}-\tau_{bc}\right)S_{ad}+\left(\theta_{bd}-\tau_{bd}\right)S_{ac}\right)\label{eq:op2-tab-mult}\\
 & = & -2i\sqrt{p_{*}^{2}}\left[P_{14}\left(2\right)_{ab,cd}-P_{41}\left(2\right)_{ab,cd}\right].\nonumber \end{eqnarray}
One can also show that other two key operators, $\left(\tau S\right)_{ab,cd}^{2}$
and $\left(\theta\tau S\right)_{ab,cd}^{2}$, can be written either
in terms of $\theta$, $\tau$ and $S$ or $P_{ij}\left(2\right)$,
according to the convenience:\begin{eqnarray}
\left(\tau S\right)_{ab,cd}^{2} & = & p_{*}^{2}\left[\frac{1}{2}\left(\theta_{ac}\tau_{bd}+\theta_{ad}\tau_{bc}+\tau_{ac}\theta_{bd}+\tau_{ad}\theta_{bc}\right)-\left(\tau_{ac}\tau_{bd}+\tau_{ad}\tau_{bc}\right)\right]\label{eq:op3-tab-mult}\\
 & = & p_{*}^{2}\left[P_{22}\left(2\right)_{ab,cd}+P_{33}\left(2\right)_{ab,cd}\right],\nonumber \\
\left(\theta\tau S\right)_{ab,cd}^{2} & = & p_{*}^{2}\left[\left(\theta_{ac}-\tau_{ac}\right)\left(\theta_{bd}-\tau_{bd}\right)+\left(\theta_{ad}-\tau_{ad}\right)\left(\theta_{bc}-\tau_{bc}\right)\right]-\left(S_{ac}S_{bd}+S_{ad}S_{bc}\right)\label{eq:op4-tab-mult}\\
 & = & 4p_{*}^{2}\left[P_{11}\left(2\right)_{ab,cd}+P_{44}\left(2\right)_{ab,cd}\right].\nonumber \end{eqnarray}

With the operators \eqref{eq:op1-tab-mult}-\eqref{eq:op4-tab-mult},
one can show, by direct inspection, that the following multiplication
table is fulfilled:

\begin{table}[H]
\centering{}\begin{tabular}{|c|c|c|c|c|c|}
\hline 
 & $P\left(2\right)_{cd,ef}$ & $\left(\tau S\right)_{cd,ef}$ & $\left(\tau S\right)_{cd,ef}^{2}$ & $\left(\theta\tau S\right)_{cd,ef}$ & $\left(\theta\tau S\right)_{cd,ef}^{2}$\tabularnewline
\hline
\hline 
$P\left(2\right)_{ab,cd}$ & $P\left(2\right)$ & $\tau S$ & $\left(\tau S\right)^{2}$ & $\theta\tau S$ & $\left(\theta\tau S\right)^{2}$\tabularnewline
\hline 
$\left(\tau S\right)_{ab,cd}$ & $\tau S$ & $\left(\tau S\right)^{2}$ & $p_{*}^{2}\left(\tau S\right)$ & 0 & 0\tabularnewline
\hline 
$\left(\tau S\right)_{ab,cd}^{2}$ & $\left(\tau S\right)^{2}$ & $p_{*}^{2}\left(\tau S\right)$ & $p_{*}^{2}\left(\tau S\right)^{2}$ & 0 & 0\tabularnewline
\hline 
$\left(\theta\tau S\right)_{ab,cd}$ & $\theta\tau S$ & 0 & 0 & $\left(\theta\tau S\right)^{2}$ & $4p_{*}^{2}\left(\theta\tau S\right)$\tabularnewline
\hline 
$\left(\theta\tau S\right)_{ab,cd}^{2}$ & $\left(\theta\tau S\right)^{2}$ & 0 & 0 & $4p_{*}^{2}\left(\theta\tau S\right)$ & $4p_{*}^{2}\left(\theta\tau S\right)^{2}$\tabularnewline
\hline
\end{tabular}\caption{Multiplication table of spin-$2$ operators\label{tab:Multiplication-table-of}}

\end{table}

Restricting the discussion for the spin-$2$ sector, the problem of
calculating the propagator is reduced to the problem of solving the
linear problem

\begin{eqnarray}
\left(xP\left(2\right)+y\tau S+z\left(\tau S\right)^{2}+u\theta\tau S+v\left(\theta\tau S\right)^{2}\right)\left(aP\left(2\right)+bS\left(2\right)\right) & \equiv & P\left(2\right),\label{eq:-1}\end{eqnarray}
for which the table above is very helpful. The general solution is
given by:

\begin{equation}
x=\frac{1}{a},\ y=\frac{b}{b^{2}p_{*}^{2}-a^{2}},\ z=-\frac{1}{a}\frac{b^{2}}{b^{2}p_{*}^{2}-a^{2}},\ u=\frac{b}{4b^{2}p_{*}^{2}-a^{2}},\ v=-\frac{1}{a}\frac{b^{2}}{4b^{2}p_{*}^{2}-a^{2}}.\label{eq:-2}\end{equation}
For the wave operator \eqref{eq:waveOperator}, $a=\frac{1}{2}p^{2}\left(\alpha+\beta p^{2}\right)$
and $b=\frac{1}{4}\left(\mu p^{2}+\lambda p^{4}\right)$, so it can
be verified that it coincides with the result obtained with the orthogonal
basis of operators as done in Sec. \ref{sec:Attainment-of-the}.

A first remark that must be made is that, without defining an orthonormal
basis of operators, there are some ambiguities for choosing the fundamental
blocks that close the algebra of operators, such as the one in table
\ref{tab:Multiplication-table-of}. In this way, such a procedure
to obtain the propagator may yield redundant operators that harm the
task of inverting the wave operator. Another remark is that, even
with the propagator at hand, the use of a non-orthonormal basis also
renders difficult the physical interpretation of the propagating modes,
since, as discussed in the Sec. \ref{sec:Spectral-Consistency-Analysis},
the degree-of-freedom basis of operators allows the splitting of the
propagator into independent sectors that result in a direct identification
of the spins of the propagating particles.

\end{document}